\documentclass[letterpaper, 10 pt, conference]{ieeeconf}   
\IEEEoverridecommandlockouts                    
\let\labelindent\relax
\usepackage{enumitem}

\usepackage{mathtools,amsthm}

\usepackage{graphics} 
\usepackage{epsfig}
\usepackage{times} 
\usepackage{amsmath,amssymb,amsfonts}
\usepackage{mathrsfs}
\usepackage{algorithmic}
\usepackage{graphicx}
\usepackage{textcomp}
\usepackage{varioref}
\usepackage{import}
\usepackage{framed,xcolor}
\usepackage{color}
\usepackage{comment}
\usepackage{multirow}
\usepackage{epstopdf}   
\usepackage{psfrag}
\usepackage{breqn}
\usepackage{float}
\usepackage{mathtools}
\usepackage{booktabs}
\usepackage{soul}
\usepackage{amsbsy}
\usepackage[bottom]{footmisc}
\usepackage{lineno}
\usepackage{cleveref}
\usepackage{microtype}
\usepackage{comment}

\definecolor{arash}{rgb}{0.8,0.8,1}
\definecolor{seb}{rgb}{0.8,1,0.8}
\definecolor{seb2}{rgb}{0.5,.5,1}
\definecolor{arash2}{rgb}{0,.5,0}
\definecolor{wenqi}{rgb}{1,.75,0.79}
\definecolor{wenqi2}{rgb}{1,.75,0.79}

\newcommand{\vect}[1]{\ensuremath{\boldsymbol{\mathrm{#1}}}}
\usepackage{graphicx}
\usepackage{tabularx,booktabs}
\usepackage{color}
\usepackage{multicol}
\usepackage{amsmath,amssymb}
\usepackage{amsfonts}
\usepackage{textcomp}
\usepackage{psfrag}
\usepackage{multimedia}
\usepackage{fancybox}
\usepackage{comment}
\usepackage{soul}
\makeatletter
\newcommand{\biggg}{\bBigg@{1.6}}  

\makeatother

\definecolor{seb}{rgb}{0.8,1,0.8}
\definecolor{arash}{rgb}{0.8,0.8,1}

\newcounter{lastnote}

\title{\LARGE 
Multi-agent Battery Storage Management using MPC-based Reinforcement Learning}
\author{Arash Bahari Kordabad, Wenqi Cai, Sebastien Gros
\thanks{The authors are with Department of Engineering Cybernetics, Norwegian University of Science and Technology (NTNU), Trondheim, Norway. E-mail:{\tt\small\{Arash.b.kordabad, wenqi.cai, sebastien.gros\}@ntnu.no}}
}
\begin{document}
\bstctlcite{IEEEexample:BSTcontrol}

\maketitle
\thispagestyle{empty}
\pagestyle{empty}
\begin{abstract}
In this paper, we present the use of Model Predictive Control (MPC) based on Reinforcement Learning (RL) to find the optimal policy for a multi-agent battery storage system. A time-varying prediction of the power price and production-demand uncertainty are considered. We focus on optimizing an economic objective cost while avoiding very low or very high state of charge, which can damage the battery. We consider the bounded power provided by the main grid and the constraints on the power input and state of each agent. A parametrized MPC-scheme is used as a function approximator for the deterministic policy gradient method and RL optimizes the closed-loop performance by updating the parameters. Simulation results demonstrate that the proposed method is able to tackle the constraints and deliver the optimal policy.
\end{abstract}
\section{INTRODUCTION}
\par Increasingly many electricity consumers actively participate in the power system through bidirectional power trades~\cite{lee2015coordinated}. In order to improve the efficiency of power transmission and the power quality, one of the key technologies is based on the Energy Storage Systems (ESS)~\cite{rastler2010electricity}. A multi-agent battery storage system, usually includes several batteries that are connected to a main grid. The main grid exchanges the power with all of the batteries and the batteries attempt to optimize their own cost. Since the total power exchanged by the main grid is limited at each time, finding an optimal policy that satisfies this restriction is challenging.
\par Making decisions for the power system to optimize an economic cost in the presence of different forms of uncertainties is the object of recent publications~\cite{gross2020stochastic,gross2020using}. In smart grids, the uncertainties mainly arise from the imperfect forecasts of the long-term prices and the power production-demand. Reinforcement Learning (RL) offers tools for tackling Markov Decision Processes (MDP) without having an accurate knowledge of the probability distribution underlying the state transition~\cite{sutton2018reinforcement,bertsekas2019reinforcement}. RL seeks to optimize the parameters underlying a given policy in view of minimizing the expected sum of a given stage cost. RL methods are usually either directly based on an approximation of the optimal policy or indirectly based on an approximation of the action-value function. Policy gradient methods directly attempt to find the optimal policy parameters by optimizing the closed-loop performance. Q-learning and Least Squares Temporal Different (LSTD) are among the algorithms that capture the action-value function~\cite{lagoudakis2003least}. Regarding the approximation of the generic optimal policy and optimal action-value function, Fuzzy Neural Network and Deep Neural Networks (DNNs) are common choices~\cite{ELearning}. 
\par In the smart grids context, usually there are reasonable forecasts of the statistics of the uncertainties and a knowledge of the systems dynamics. Therefore, using a structured function approximation such as Model Predictive Control (MPC) scheme can be beneficial. Indeed, MPC uses the predicted information and model to provide a reasonable but usually suboptimal policy~\cite{MPCbook}. Moreover, MPC is able to handle the high-dimensionality of the forecasts. In~\cite{gros2019data}, it is shown that adjusting the model, cost, and constraints of the MPC could achieve the best closed-loop performance, and RL is proposed as a possible approach to perform that adjustment in practice. Recent researches have developed further the combination of RL and MPC (see e.g.~\cite{kordabad2021mpc,nejatbakhsh2021reinforcement,bahari2021reinforcement,koller2018learning,bahari2021verification}). 
\par In this paper, considering the time-varying prediction of the spot market and the production-demand uncertainty, we use an MPC-scheme to minimize the running cost of the system, while penalizing extreme State-of-Charge (SOC). A low-level controller monitors the SOC in real time and prevents violating the constraints by buying or selling more power if needed~\cite{leao2010lead}. We suppose that all the agents are connected to a main grid, and each battery stores or releases a limited amount of power at every time instant. The deterministic policy gradient method and the LSTD method are adopted to update the policy parameters and action-value parameters, respectively. The simulation results show that our proposed MPC-based RL method is capable of finding the optimal MPC parameters for the multi-agent battery storage system.
\par The rest of the paper is structured as follows. Section \ref{sec:Model} provides the multi-agent battery storage dynamics and details the economic objective and constraints of the problem. Section \ref{sec:MPC-PG} formulates the centralized MPC-scheme method via the MPC-based policy and it presents the policy gradient method that used to find the optimal policy. Section \ref{sec:sim} presents the simulations and section \ref{sec:conc} delivers a conclusion.
\section{Problem Formulation}\label{sec:Model}
In this section, we formulate the battery storage dynamics, the economic objective function with state constraints for a multi-agent system, and peak power constraints over time. 
\subsection{Dynamics}
\begin{figure}[ht!]
\centering
\includegraphics[width=0.48\textwidth]{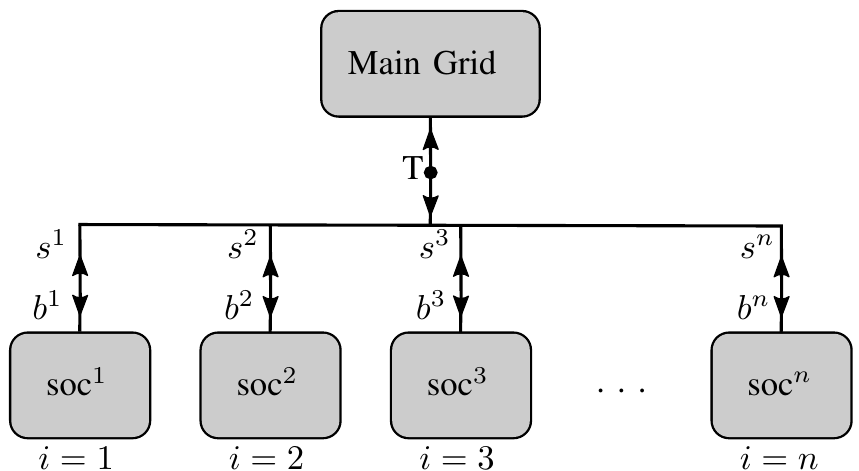}
	\caption{Multi-agent battery storage system}
	\label{fig:0}
\end{figure}

Photovoltaic (PV) battery systems allow households to participate in a more sustainable energy system (\cite{gross2020stochastic}). The battery storage dynamics can be written as the following linear system:
\begin{align}
\label{eq:Dyna0}
\mathrm{soc}^i_{k+1} &= \mathrm{soc}^i_{k} + \alpha^i\left(\Delta^i_k + b^i_k- s^i_k \right), 
\end{align}
where $i\in[1,\ldots,n]$ is the $i^{\mathrm{th}}$ battery, $n$ is the number of batteries, subscript $k=0,1,\ldots$ denotes the physical time, $\mathrm{soc}^i_k\in[0,1]$ is the State-of-Charge (SOC) of the battery and the interval $[0,1]$ represents the SOC levels considered as non-damaging for the battery (typically 20\%-80\% range of the physical SOC). Constant $\alpha^i$ is a positive value that reflects the battery size. Process noise $\Delta^i_k\sim\mathcal N\left(\bar \delta^i, \sigma^i\right)$ is the difference between the local power production-demand over the sampling time interval $[k,k+1]$, which--for the sake of simplicity--is considered as a Normal centred random variable, where $\bar\delta^i$ and $\sigma^i$ are the mean and variance of the Gaussian distribution. Input $b^i_k$($s^i_k$) $\in[0,\bar U^i]$ is the average power bought (sold) from (to) the power grid over time interval $[k,k+1]$, where $\bar U^i$ is the bound for the buying (selling) energy for the $i^{\mathrm{th}}$ battery. Fig. \ref{fig:0} illustrates the multi-agent battery system, where the batteries are connected to the main grid at point $T$.
\subsection{Objective Function}
\par Economic costs for smart grids are usually linear, based on the difference between the profit made by selling electricity to the power grid, and the losses incurred from buying it~(see e.g., \cite{harsha2014optimal}). Hence, each battery has the following economic stage cost:
\begin{align}
\label{eq:L}
L(b^i_k,s^i_k) = \phi^i_b b^i_k-\phi^i_s s^i_k,
\end{align}
where $\phi^i_b \geq 0$ and $\phi^i_s\geq 0$ are the (time-varying) buying and selling prices, respectively. 
\par In the context of RL, we seek a control policy $\vect\pi$ that maps the state space to the input space and minimizes a closed-loop performance, which can be defined as an infinite-horizon expected sum stage costs. For the battery storage dynamics \eqref{eq:Dyna0} with stage cost \eqref{eq:L} and constraint $\mathrm{soc}^i\in[0, 1]$ for all $1 \leq i\leq n$, the modified stage cost $\tilde {L}$ for the centralized system can be defined as follows:
\begin{align}
    \tilde {L}(\vect s_k,\vect a_k)= \sum_{i=1}^n \Big(& L(b_k^i,s_k^i)+p^i\max(\mathrm{soc}_k^i-0.9,0)\nonumber\\&+p^i\max(0.1-\mathrm{soc}_k^i,0)\Big),
\end{align}
where $p^i$ is a large constant that penalizes the state constraints within $10\%$ of the bound $\mathrm{soc}^i_k\in[0,1]$. Vectors $\vect s_k=\mathrm{soc}_k^{1,\ldots,n}$ and $\vect a_k=\{b_k^1-s_k^1,\ldots,b_k^n-s_k^n\}$ describe the entire system states and inputs vectors, respectively. A very low or very high state of charge decrease the battery lifetime~\cite{wikner2018extending}. Note that under optimality condition, the buying and selling variables can not be non-zero at the same time, then the difference of buying and selling $b_k^i-s_k^i$ can be considered as the input of the system (\cite{kordabad2021mpc}). The closed-loop performance $J$ reads as:
\begin{align}
\label{eq:J}
J(\vect\pi)=\mathbb E_{\vect\pi}\Bigg[&\sum_{k=0}^\infty \gamma^k \tilde {L}(\vect s_k,\vect a_k) \Bigg|\vect a_k=\vect \pi(\vect s_k)\Bigg],
\end{align}
where $\gamma\in(0,1]$ is the discount factor and expectation $\mathbb E_{\vect\pi}$ is taken over the distribution of the Markov chain in closed-loop with policy $\vect\pi$. The performance for agent $i^{\mathrm{th}}$ is then defined as:
\begin{align}
\label{eq:J_i}
J^i(\vect\pi)=\mathbb E_{\vect\pi}&\Bigg[\sum_{k=0}^\infty \gamma^k\Big(  L(b^i_k,s^i_k)+p^i\max(\mathrm{soc}_k^i-0.9,0)\nonumber\\&+p^i\max(0.1-\mathrm{soc}_k^i,0)\Big)\Bigg|\vect a_k=\vect \pi(\vect s_k)\Bigg].
\end{align}
\subsection{Peak Power Constraint at point $T$}
\par Electricity customers usually have different power demands during the day. In a multi-agent battery problem with a common main grid, optimizing the power peaks is critical. The methods for flattening the load curve are often called \emph{peak shaving}. In order to formulate peak power constraints at point $T$ (see Fig.\ref{fig:0}), we first define $P_k$ as the maximum power amount exchanged with the main grid, i.e.:
  \begin{align}
     P_{k}&=\max\left(\sum _{i=1}^n b^i_{k},\sum _{i=1}^n s^i_{k}\right).
 \end{align}
Assume that $P_{k}$ is restricted by the following upper bound over time:
\begin{align}\label{eq:ppeak}
P_{k}\leq \bar P \,, \quad \forall k\geq0 , 
\end{align}
where $\bar P>0$ is the maximum allowed power amount that can be exchanged with the main grid. The maximum grid power $\bar P$ is assumed to be less than sum of the maximum exchanged power for the each battery. i.e.:
\begin{align}
\bar P < \sum _{i=1}^n \bar U^i \,.
\end{align}
Otherwise \eqref{eq:ppeak} holds by construction. 
\par Next section details the parametrization of the MPC-scheme that is used as an approximator for the RL method and provides the policy gradient formulation to update the parameters.
\section{MPC-based Deterministic Policy Gradient}\label{sec:MPC-PG}
Using MPC to support the approximations of the value function, the action-value function, and the policy has been proposed and justified in~\cite{gros2019data}. In this section, we detail this approach. We utilize the deterministic policy gradient method to adjust the MPC parameters and improve the closed-loop performance.
\subsection{Centralized MPC-scheme}
We focus on an MPC-based approximation of the optimal policy. RL is used to adjust the parameters $\vect \theta$ in the MPC-scheme to handle model uncertainties and the process noise $\Delta^i$. Furthermore, RL will tune the parameters so as to push the SOC to a safe region ($10\%-90\%$ of the state of the charge). Note that the outside the interval $[0.1, 0.9]$ for $\mathrm{soc}$, even if it is feasible, but it may damage the battery and reduce its lifetime. In order to provide a MPC-based policy approximator for RL, consider the following MPC scheme parameterized by $\vect \theta$:
\begin{subequations}\label{eq:MPC}
\begin{align}
    &\min_{\hat {\mathrm{soc}}, \hat b,\hat s, \vect\sigma}\,\,\,\, \sum_{i=1}^{n}\Bigg( {\vect\omega_f^i}^\top\vect\sigma^i_N+ T_{\vect\theta}(\hat{\mathrm{soc}}^i_{N})+\label{eq:MPC:cost} \\&\null\qquad\qquad\qquad \sum_{j=0}^{N-1}\gamma^j\left(  L_{\vect\theta}(\hat{b}^i_j,
    \hat s^i_j)+\phi_{\vect\theta}(\hat{\mathrm{soc}}^i_{j})+{\vect\omega^i}^\top\vect\sigma^i_j\right)\Bigg)\nonumber \\
    &\quad\quad\mathrm{s.t.}\quad
    \forall i=1,\ldots,n , \quad\forall j=0,\ldots,N-1\nonumber
    \\&\null \,\,\qquad\qquad \hat{\mathrm{soc}}^i_{j+1} = \hat{\mathrm{soc}}^i_{j} + \theta_{\alpha}^i(\hat b^i_j-\hat s^i_j)+\theta_{\delta}^i, \label{eq:MPC:dynamics}\\
    &\null \,\,\qquad\qquad [\hat{\mathrm{soc}}^i_{j}-0.9,0.1-\hat{\mathrm{soc}}^i_{j}]^\top\leq \vect\sigma^i_j,0\leq\vect\sigma^i_j \label{eq:MPC:state:cons1}\\&\null \,\,\qquad\qquad[\hat{\mathrm{soc}}^i_{N}-0.9,0.1-\hat{\mathrm{soc}}^i_{N}]^\top\leq \vect\sigma^i_N, 0\leq\vect\sigma^i_N\label{eq:MPC:state:cons2}\\
    &\null\,\,\qquad\qquad 0\leq \hat b^i_j\leq \bar U^i,\qquad
0\leq \hat s^i_j\leq \bar U^i,\label{eq:MPC:input:cons}\\ &\null\,\,\qquad\qquad \sum _{i=1}^n \hat b^i_j \leq \bar P, \,\,\qquad \sum _{i=1}^n \hat s^i_j \leq \bar P,\label{eq:MPC:ppeak1}\\&\null\,\,\qquad\qquad \hat{\mathrm{soc}}^i _{0}=\mathrm{soc}^i _{k}, \label{eq:MPC:equality}
\end{align}
\end{subequations}
where $\hat {\mathrm{soc}}=\hat{\mathrm{soc}}^{1,\ldots,n}_{0,\ldots,N}$, $\hat b=\hat b^{1,\ldots,n}_{0,\ldots,N-1}$, $\hat s=\hat s^{1,\ldots,n}_{0,\ldots,N-1}$, $\vect\sigma=\vect\sigma^{1,\ldots,n}_{0\ldots,N}$ are the primal decision variables for the predicted state, buying, selling and slacks, respectively. Subscript $j$ is the MPC prediction step and $N$ is the horizon length. We relax the stage and terminal state inequalities by the positive slack variables $\vect\sigma^i_j$ and $\vect\sigma^i_N$, and penalize them by positive constant weights $\vect \omega^i$  and $\vect \omega_f^i$, respectively. This prevents the infeasibility of the MPC in the presence of the process noise in the real system \eqref{eq:Dyna0} out of the interval $[0.1, 0.9]$ for the states. Stage cost $\phi_{\vect\theta}$ and terminal cost $T_{\vect\theta}$ are the additional parametric costs, depending on the states that allows the MPC-scheme \eqref{eq:MPC} to provide a more generic function approximator. Moreover, because of the stochasticity of the real system and the existence of different uncertainties in the system, we select the parameterized economic cost $L_{\vect\theta}$ as a generic function different with the true $L$ in \eqref{eq:L} and let RL to adjust its parameters. Parameters
 $\theta^i_{\alpha}$ and $\theta^i_{\delta}$, among the adjustable parameters $\vect\theta$, are dedicated to capture the model correction. We summarize \eqref{eq:MPC} as follows:
 \begin{itemize}
     \item Cost \eqref{eq:MPC:cost} includes the discounted economic cost $L_{\vect\theta}$, additional stage cost $\phi_{\vect\theta}$ and terminal cost $T_{\vect\theta}$ and penalty for the slack variables $\vect\sigma^i_j$ and $\vect\sigma^i_N$.
     \item Equality constraint \eqref{eq:MPC:dynamics} represents the parameterization for the deterministic model of the real system \eqref{eq:Dyna0}.
     \item Inequality constraints \eqref{eq:MPC:state:cons1} and \eqref{eq:MPC:state:cons2} are the relaxed state constraints with positive slacks for each battery.
     \item Inequality constraint \eqref{eq:MPC:input:cons} are the input constraints for each battery.
     \item Inequality constraints \eqref{eq:MPC:ppeak1} represent the power peak constraint for the grid.
     \item Equality constraint \eqref{eq:MPC:equality} initializes the MPC-scheme at current state $\mathrm{soc}_k^i$.
 \end{itemize}
The parameterized deterministic policy for agent $i$ at time $k$ can be obtained as: 
\begin{align}
     \pi_{\vect\theta}^{i}(\vect s_k)= \hat b_0^{i\star}(\vect s_k,\vect \theta)- \hat s_0^{i\star}(\vect s_k,\vect \theta),
\end{align}
 where $\hat b_0^{i\star}$ and $\hat s_0^{i\star}$ are the first elements of $\hat b^{i\star}$ and $\hat s^{i\star}$, which are the solutions of the MPC scheme \eqref{eq:MPC} associated to the decision variables $\hat b^{i}$ and $\hat s^{i}$. Then the parametric centralized policy extracted from the MPC-scheme \eqref{eq:MPC} is written as follows:
 \begin{align}\label{eq:pi}
  \vect\pi_{\vect\theta}(\vect s_k)=[\pi_{\vect\theta}^1,\ldots,\pi_{\vect\theta}^n]^\top   
 \end{align}
The input $\vect a_k$ is selected according to the corresponding parametric policy $\vect\pi_{\vect\theta}$ in \eqref{eq:pi} with possible addition of small random exploration.
\subsection{Low-level Control}
In the smart grid context, there is usually a low-level control that monitors the current state of charge (which has a $1$h sampling time) and power demand/production. If the states tend to violate the constraints $\mathrm{soc}^i_k\in[0,1]$, the low-level control (which works at a lower sampling time, e.g., every second) would decide to buy or sell more power to keep the states in the feasible interval \cite{leao2010lead,avila2019state}.
\subsection{Deterministic Policy Gradient Method}
The deterministic policy gradient method optimizes the policy parameters directly via gradient descent steps on the performance function $J$, defined in \eqref{eq:J}. The update rule is as follows:
\begin{align}
    \vect\theta \leftarrow \vect\theta-\alpha  \nabla _{\vect\theta}J(\vect\pi _{\vect\theta}),
\end{align}
where $\alpha>0$ is the step size. Applying the deterministic policy gradient method, developed by \cite{silver2014deterministic}, the gradient of $J$ with respect to parameters $\vect\theta$ is obtained as:
\begin{align}\label{eq:dj}
    \nabla _{\vect\theta}J(\vect\pi _{\vect\theta}) = \mathbb E\left[{\nabla _{\vect\theta} }{\vect\pi _{\vect\theta} }(\vect s){\nabla _{\vect a}}{A_{{\vect\pi _{\vect\theta} }}}(\vect s,\vect a)|_{\vect a=\vect \pi _{\vect\theta}}\right],
\end{align}
where $A_{{\vect{ \pi}}_{\vect{\theta}}}(\vect s,\vect a)=Q_{\vect{\pi}_{\vect{\theta}}}(\vect s,\vect a) - V_{\vect{\pi}_{\vect{\theta}}}(\vect s)$ is the \textit{advantage function} associated to  ${\vect{ \pi}}_{\vect{\theta}}$, and where $Q_{\vect{\pi}_{\vect{\theta}}}$ and $V_{\vect{\pi}_{\vect{\theta}}}$ are the action-value function and value function of the policy ${\vect{ \pi}}_{\vect{\theta}}$, respectively, defined as follows:
\begin{subequations}
\begin{align}
{Q_{\vect\pi _{\vect\theta }}}\left( {\vect s,\vect a} \right) &= \tilde L\left( {\vect s,\vect a} \right) + \gamma {\mathbb E}\left[ {{V_{\vect\pi _{\vect\theta }}}\left( {{\vect s^ + }|\vect s,\vect a} \right)} \right] \label{eq:Q}\\
{V_{\vect\pi _{\vect\theta }}}(\vect s) &= {Q_{\vect\pi _{\vect\theta }}}\left( \vect s,{\vect\pi _{\vect\theta }\left( \vect s \right)} \right), \label{eq:V}
\end{align}
\end{subequations}
where $\vect s^+$ is the subsequent state of the state-input pair ($\vect s, \vect a$). Under some conditions~\cite{silver2014deterministic},  the action-value function $Q_{{\vect\pi _{\vect\theta} }}$ in \eqref{eq:dj} can be replaced by an approximator ${Q_{\vect w}}$ without affecting the policy gradient. Such an approximation is labelled \textit{compatible} and can, e.g., take the form:
\begin{align}
\label{eq:Q_w}
    Q_{\vect w}\left(\vect s,\vect a\right ) = \left(\vect a - {\vect \pi _{\vect\theta} }\left(\vect s\right )\right )^{\top}\nabla _{\vect\theta}\vect \pi _{\vect\theta}\left(\vect s\right )^{\top}\vect w + V^{\vect v}\left(\vect s\right ),
\end{align}
where $\vect w$ is a parameter vector estimating the action-value function $Q_{{\vect{ \pi}}_{\vect{\theta}}}$ and $V^{\vect v}\approx V_{\vect \pi_{\vect \theta}}$ is a baseline function approximating the value function. The parameterized value function $V^{\vect v}$ can, e.g., take the linear form:
\begin{align}
\label{eq:V_v}
    {V^{\vect v}} \left(\vect s\right ) =\Phi\left(\vect s \right)^\top {\vect v},
\end{align}
where $\Phi(\vect s)$ is a state feature vector and $\vect v$ is the corresponding parameter vector. The parameters $\vect w$ and $\vect v$ of the action-value function approximation \eqref{eq:Q_w} ought to be the solution of the Least Squares (LS) problem:
\begin{align}
\label{eq:error}
    \min_{\vect w, \vect v} \mathbb{E} \left[\big( Q_{\vect\pi_{\vect\theta}}(\vect s,\vect a)-Q_{\vect w} (\vect s,\vect a)\big )^2\right].
\end{align}
In this paper, the LS problem in \eqref{eq:error} is tackled via Least Squares Temporal Difference (LSTD) method (see e.g., \cite{lagoudakis2003least}) based on the stage cost $\tilde L$. LSTD belongs to \textit{batch method}, seeking to find the best fitting value function and action-value function, and it is more sample efficient than other methods.

\par The primal-dual Karush–Kuhn–Tucker (KKT) conditions underlying the MPC scheme \eqref{eq:MPC} is written as:
\begin{align}
    \vect R = {\left[ {\begin{array}{*{20}{c}}
{{\nabla _{\vect \xi}}{\mathcal L_{\vect\theta} }}&{{\vect G_{\vect\theta} }}&{\mathrm{diag}\left(\vect\mu\right) \vect H_{\vect\theta} }
\end{array}} \right]^\top},
\end{align}
where $\vect\xi=\{\hat {\mathrm{soc}}, \hat b,\hat s, \vect\sigma\}$ is the primal decision variable. Operator ``$\mathrm{diag}$" assigns the vector elements onto the diagonal position of a square matrix. $\mathcal{L}_{\vect \theta}$ is the associated Lagrange function of the MPC \eqref{eq:MPC}, written as:
\begin{align}
\mathcal{L}_{\vect \theta}(\vect y) = \Psi_{\vect \theta} + \vect\lambda^\top \vect G_\theta  + \vect\mu^\top \vect H_{\vect \theta},
\end{align}
where $\Psi_\theta$ is the MPC cost \eqref{eq:MPC:cost}, $\vect G_\theta$ gathers the equality constraints and $\vect H_\theta$ collects the inequality constraints of the MPC \eqref{eq:MPC}. Vectors $\vect\lambda,\vect\mu$ are the associated dual variables. Argument ${\vect y}$ reads as ${\vect y} =\{\vect\xi,\vect\lambda,\vect\mu\}$ and $ {\vect y}^{\star}$ refers to the solution of the MPC \eqref{eq:MPC}.  The policy sensitivity ${\nabla _{\vect \theta} }{\vect \pi _{\vect \theta} }$ required in Eq. \eqref{eq:dj} can then be obtained as follows~(\cite{gros2019data}):
\begin{align}
\label{eq:sensetivity}
{\nabla _{\vect \theta} }{\vect \pi _{\vect \theta} }\left(\vect  s \right) =  - {\nabla _{\vect\theta} }{\vect R }\left( {\vect y^\star},\vect s,\vect\theta\right){\nabla _{\vect y}}{\vect R }{\left( {\vect y^\star},\vect s,\vect\theta \right)^{ - 1}}\frac{\partial {\vect y}}{\partial {\vect u_0}},
\end{align}
where $\vect u_0$ is the first input variable, defined as follows:
\begin{align}
    \vect u_0=[\hat b_0^{1}-\hat s_0^{1},\ldots,\hat b_0^{n}-\hat s_0^{n}]^\top.
\end{align}
Next section provides the simulation results of the proposed method for a simple configuration of the multi-agent battery storage system.
\section{Simulation}\label{sec:sim}
In this section we illustrate the simulation results of the MPC-based deterministic policy gradient method for a 3-agent battery storage problem. 
\par The state feature $\Phi(\vect s)$ for the value function approximator $V^{\vect v}(\vect s)$ in \eqref{eq:V_v} is selected as a vector of quadratic monomials as follows:
\begin{align}
   \Phi(\vect s)=
    \left[(\hat{\mathrm{soc}}^1)^2,(\hat{\mathrm{soc}}^2)^2,(\hat{\mathrm{soc}}^3)^2, 
\hat{\mathrm{soc}}^1,\hat{\mathrm{soc}}^2,\hat{\mathrm{soc}}^3,1\right]^{\top}. 
\end{align}
For the sake of simplicity, we don't consider the joint state effects in the value function.

The parameterized economic cost $L_{\vect\theta}$, additional stage cost $\phi_{\vect\theta}$, and terminal cost $T_{\vect\theta}$ in the MPC-scheme \eqref{eq:MPC} are selected as follows:
\begin{subequations}\label{eq:approx}
\begin{align}
\label{eq:L:theta}
&L_{\vect\theta}(\hat b^i_j,\hat s^i_j) = ( \phi^i_b+\theta^i_b)\hat b^i_j-(\phi^i_s+\theta^i_s) \hat s^i_j \\
&\phi_{\vect\theta}(\hat{\mathrm{soc}}^i_{j})=\phi^i_1(\hat{\mathrm{soc}}^i_{j})^2+\phi^i_2\hat{\mathrm{soc}}^i_{j}+\phi^i_3\label{eq:phi:theta}\\
&T_{\vect\theta}(\hat{\mathrm{soc}}^i_{N})=T^i_1(\hat{\mathrm{soc}}^i_{N})^2+T^i_2\hat{\mathrm{soc}}^i_{N}+T^i_3,\label{eq:T:theta}
\end{align}
\end{subequations}
where $\theta^i_b$, $\theta^i_s$, $\phi^i_{1,2,3}$, and $T^i_{1,2,3}$ are among the adjustable parameters $\vect\theta$, i.e:
\begin{align}
    \vect\theta=\{\theta^{1,\ldots,n}_\alpha,\theta^{1,\ldots,n}_\delta,\theta^{1,\ldots,n}_b,\theta^{1,\ldots,n}_s,\phi^{1,\ldots,n}_{1,2,3},T^{1,\ldots,n}_{1,2,3}\}.
\end{align}
One can use more generic function approximators in \eqref{eq:approx}, however, in \cite{kordabad2021mpc}, it shows that, for this kind of battery storage problem, quadratic parameterizations for the stage and terminal costs in the MPC-based policy approximator are rich enough to capture the optimal policy. 
\par The rest of the parameter values used in the simulation are given in Table \ref{tab:table1}. 
\begin{table}[ht!]
\caption{\label{tab:table1} Parameter values.}
\centering
\begin{tabular}{cc|cc}
\hline
Symbol & Value & Symbol & Value\\
\hline
$\gamma$ & $0.99$ &
$n$ & 3\\
Sampling time  & $1$h &
$N$  & 12 \\
$\alpha^i$  & $1/12$      &
$\Delta^i$  & $\mathcal{N}(0,0.5)$      \\
$\bar U^i$ & $1$&
$\bar P$ & $1.5$\\
$\vect \omega^i, \vect \omega_f^i$ & $[20,20]^\top$&
$p^i$ & $1000$\\
$\alpha$ & $5$e$-8$&
$\mathrm{soc}_0^i$ & $0.5$\\
\hline
\end{tabular}
\end{table}

\par
\begin{figure}[ht!]
\centering
\includegraphics[width=0.48\textwidth]{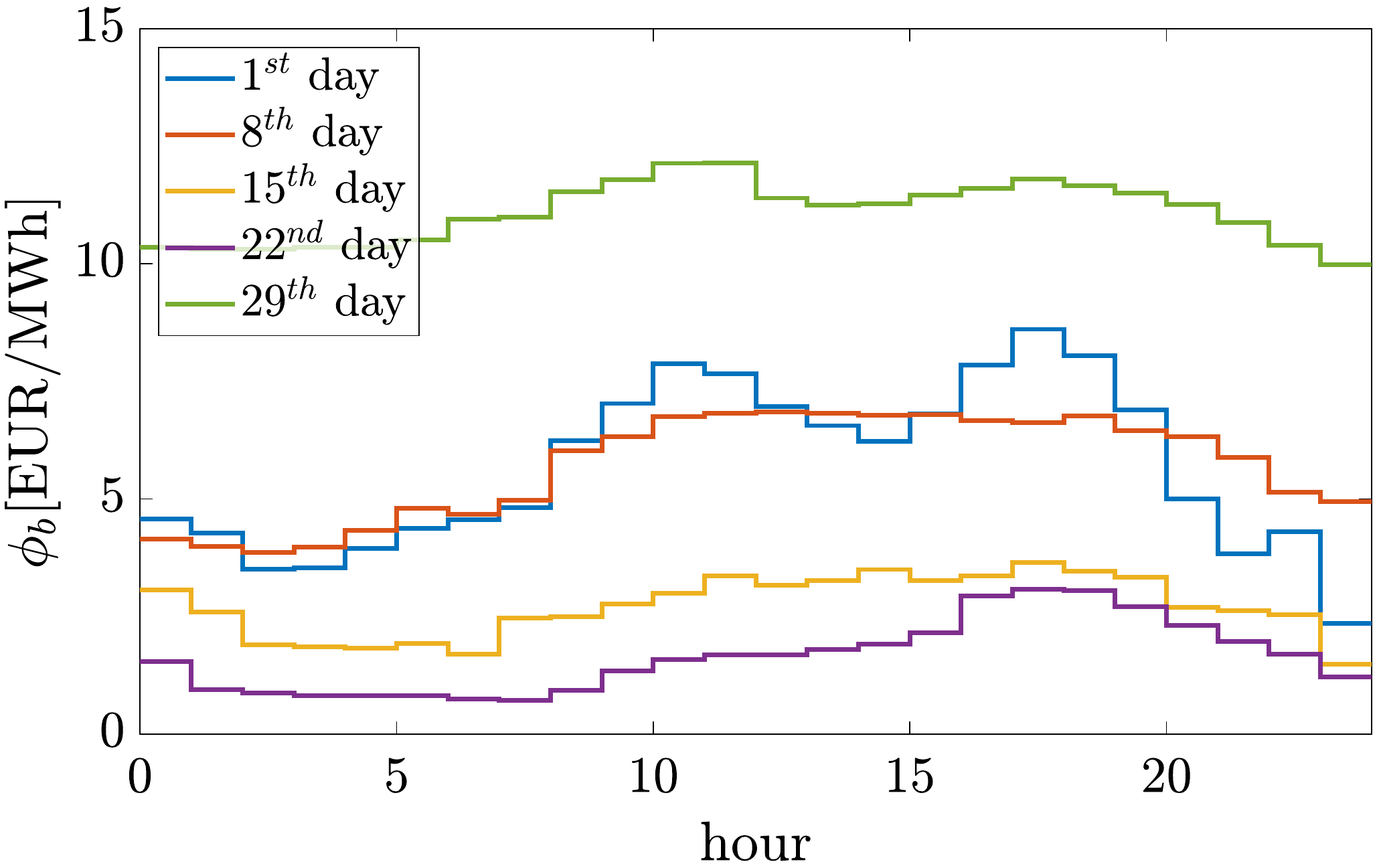}
\caption{The 24-hour buying price $\phi_b$ of Trondheim for five sampled days in Nov.2020.}
\label{fig:6}
\end{figure}

 We use the time-varying power prices of Trondheim in the simulation, which is collected from the website provided by the Nord Pool European Power Exchange~\cite{price}. Fig. \ref{fig:6} illustrates the 24-hour buying price $\phi_b$ for five sampled days of Nov.2020. For the selling price $\phi_s$, we use $\phi_s=0.5\phi_s$ at every time step. Note that the prediction horizon is selected as $N=12$, because the power prices are usually accessible for $12$-hours ahead~\cite{price}.
\par We run the simulation for $100$ months. Each month we use a repetitive $30$-days, where the states $\mathrm{soc}^i$ start from $0.6$ at the beginning of the day, and we apply the time-varying prices and consider different stochasticity for each agents. We average along $30$ days to approximate the expectations ($\mathbb{E}$) in the policy gradient \eqref{eq:dj} and LS \eqref{eq:error}, and update the parameters of the value function, action-value function, and policy at the end of each month.

\par
\begin{figure}[t!]
\centering
\includegraphics[width=0.48\textwidth]{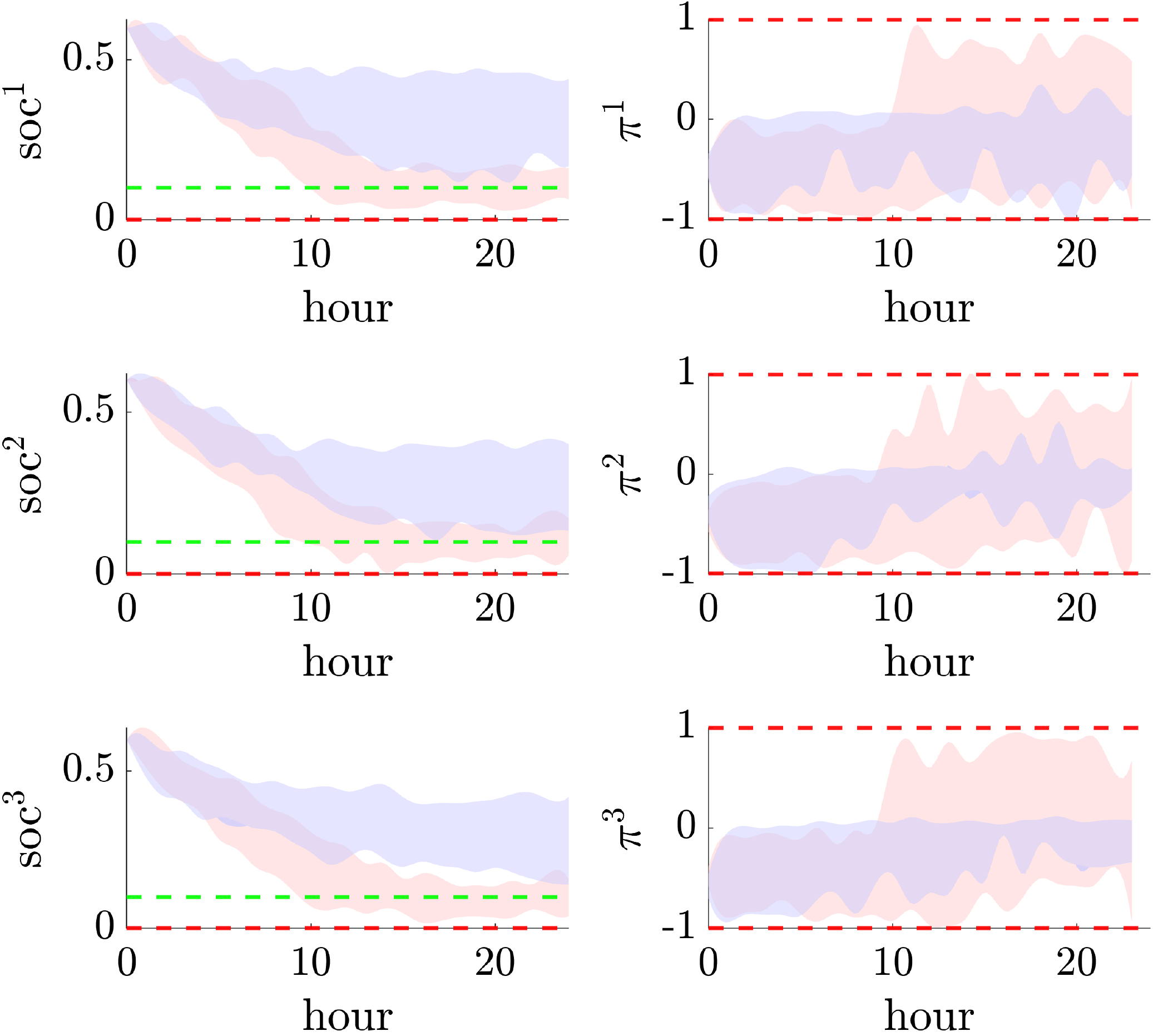}
\caption{The state and policy trajectories over time for each agent. Red: the first month of the learning, Blue: the last month of the learning.}
\label{fig:1}
\end{figure}

 Figure \ref{fig:1} shows the state and policy trajectories over time for each agent during the first and last month of the learning. The red trajectories show the states and policies for the first month. As can be seen, at the beginning month of the learning, the MPC-scheme has not learned yet and the states are sometimes in the position of lower than $10\%$ of the SOC capacity, i.e, the $\mathrm{soc}^i$ of the three agents are sometimes below $10\%$. The blue trajectories correspond to the last month of the learning. It can be seen that with learning, RL pushes the states up so as to prevent being close to the bounds of the state constraints.

\par
\begin{figure}[ht!]
\centering
\includegraphics[width=0.48\textwidth]{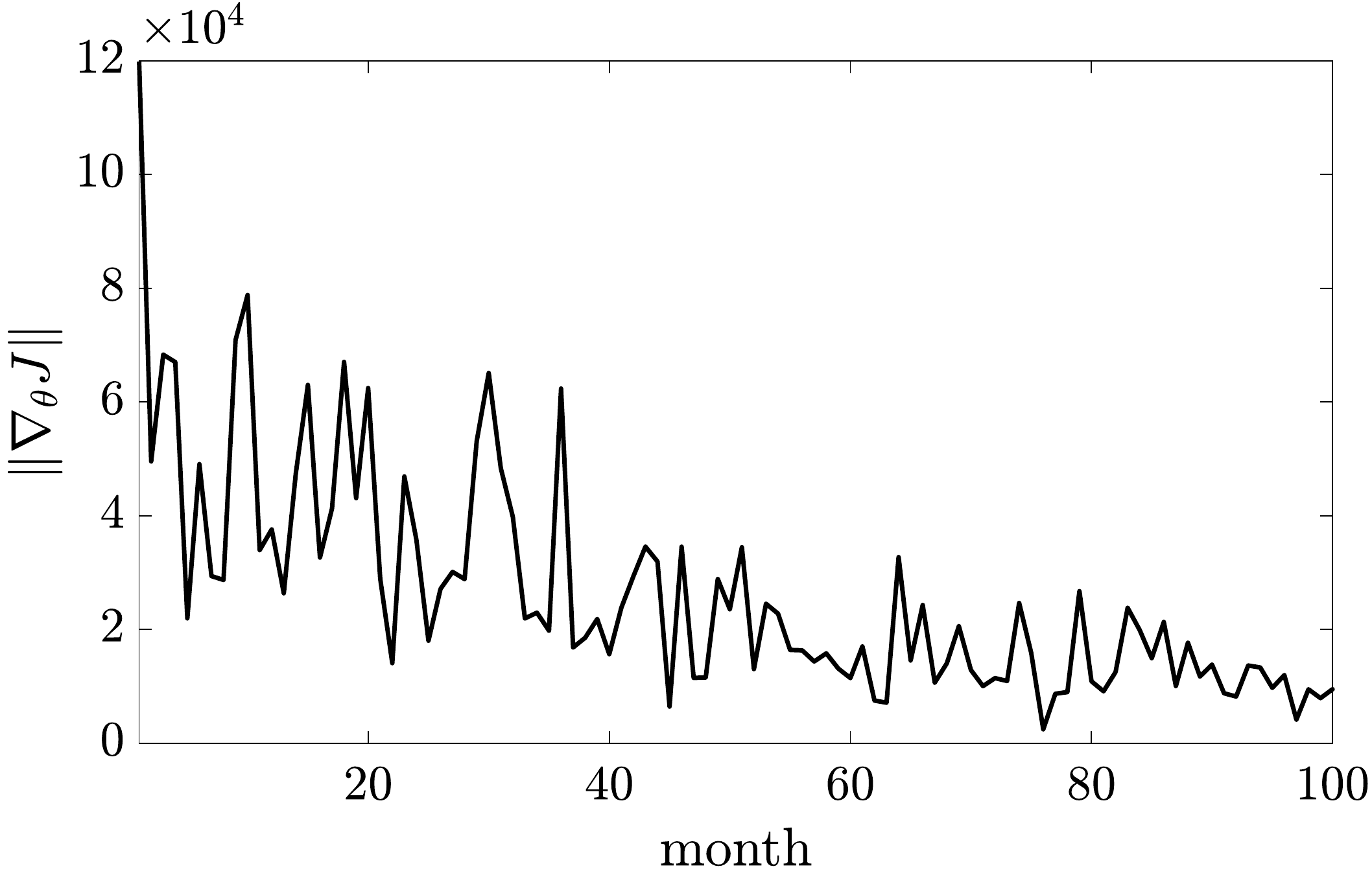}
\caption{Norm of the policy gradient $\nabla _{\vect\theta}J(\vect\pi _{\vect\theta})$ over RL-steps.}
\label{fig:3}
\end{figure}

 Figure \ref{fig:3} illustrates the norm of policy gradient $\nabla _{\vect\theta}J(\vect\pi _{\vect\theta})$ over RL-steps. Since the existence of the process noise and random exploration, the gradient is noisy, but the overall behaviour is decreasing as the parameters approach to their optimal points.
\par
\begin{figure}[ht!]
\centering
\includegraphics[width=0.48\textwidth]{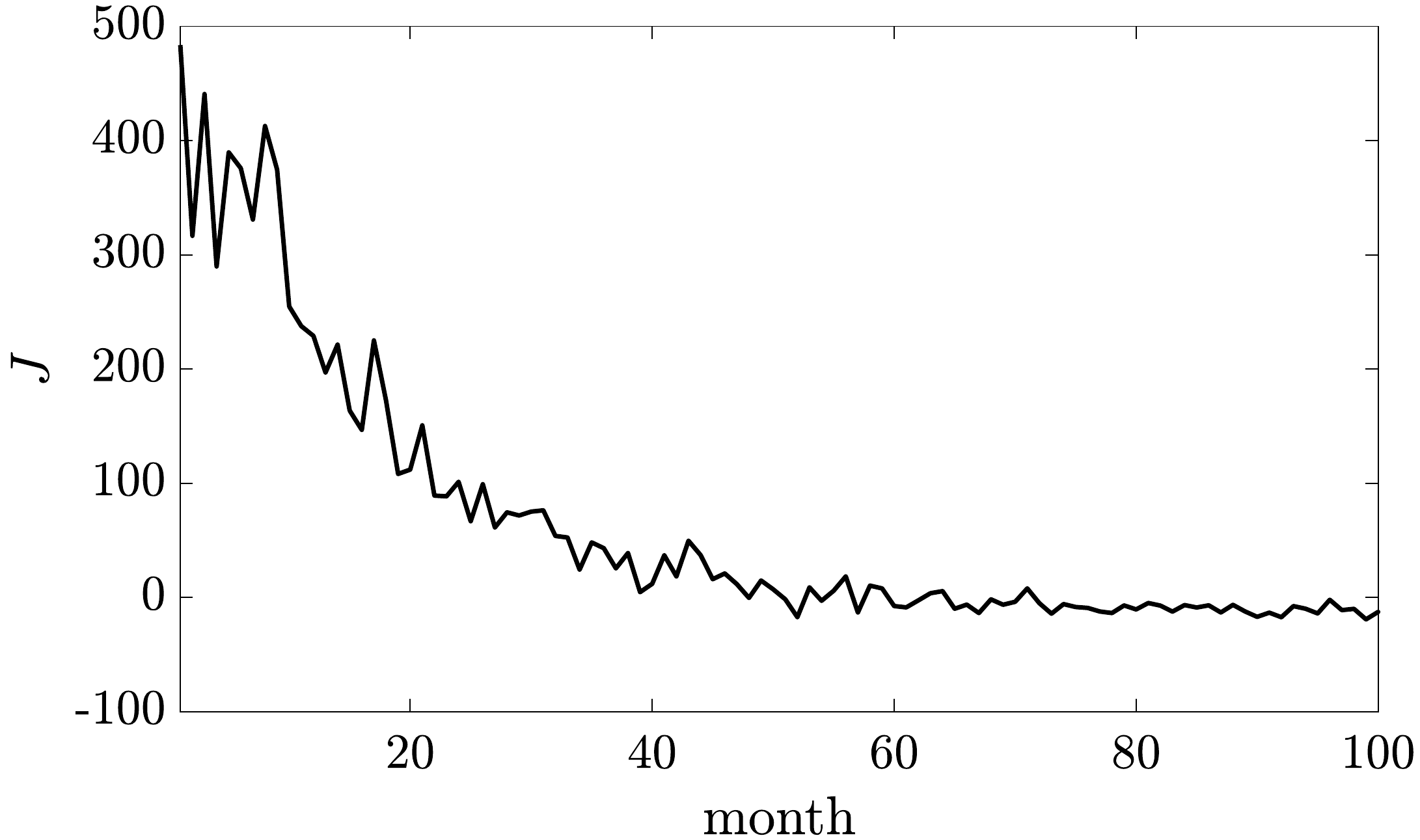}
\caption{Closed-loop performance $J$ over RL-steps.}
\label{fig:2}
\end{figure}

 The variation of the closed-loop performance $J$ is shown in Fig.\ref{fig:2}. It can be seen that the performance is improved significantly over the learning. Besides, since the value of policy gradient $\nabla _{\vect\theta}J(\vect\pi _{\vect\theta})$ is relatively large within the first twenty months, the performance $J$ drops faster in this range.

\par
\begin{figure}[ht!]
\centering
\includegraphics[width=0.48\textwidth]{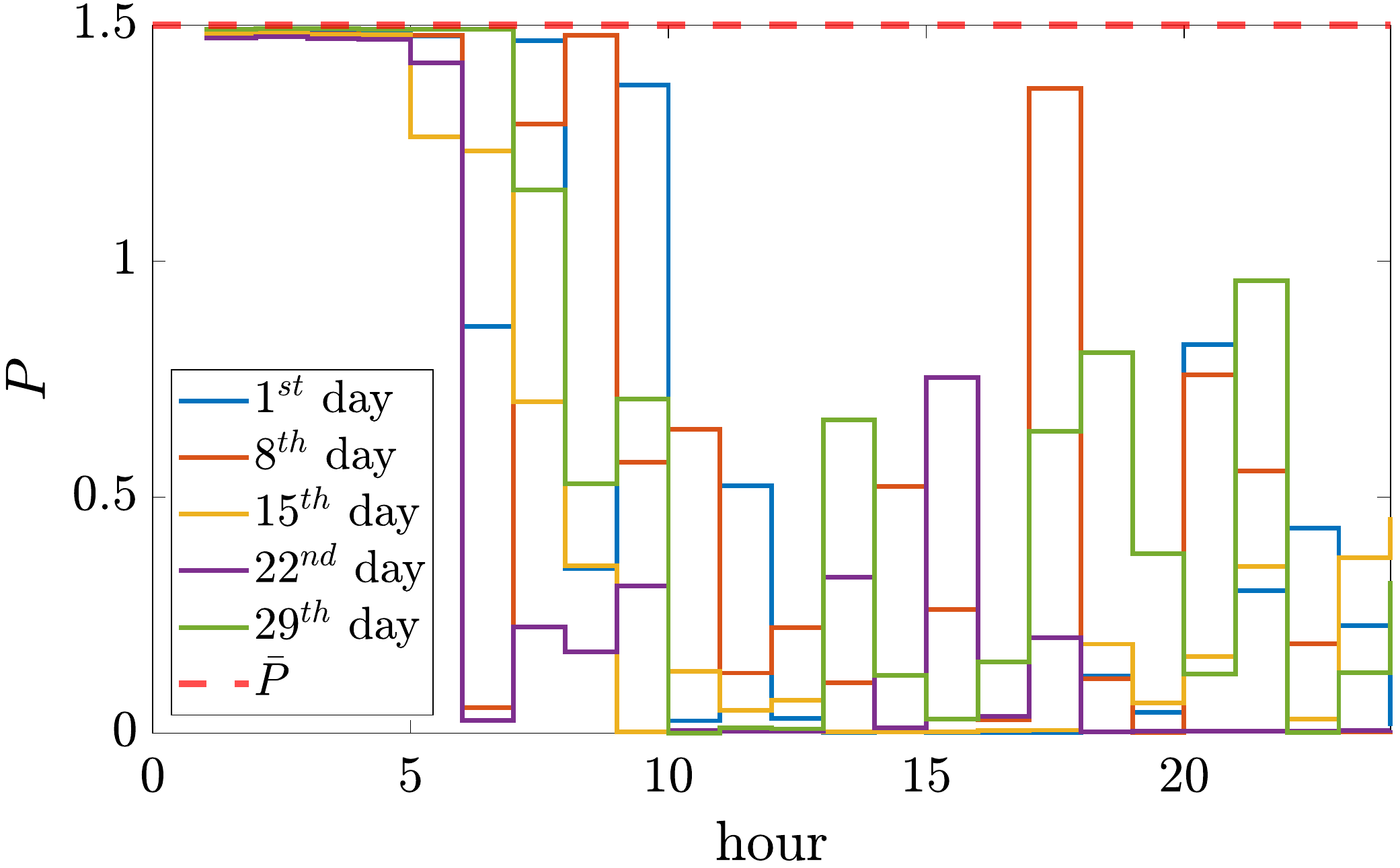}
\caption{The maximum power amount $P$  exchanged with the main grid for five sampled days.}
\label{fig:4}
\end{figure}

 Figure \ref{fig:4} presents the maximum power amount $P$ exchanged with the main grid for five sampled days in the last learning month. As can be seen, the values of $P$ comply with the upper bound constraints $\bar P$, which means the optimal policy we find can not only render the minimum economic cost for the whole system but also meet the power peak constraints on the main grid.
\par
\begin{figure}[ht!]
\centering
\includegraphics[width=0.48\textwidth]{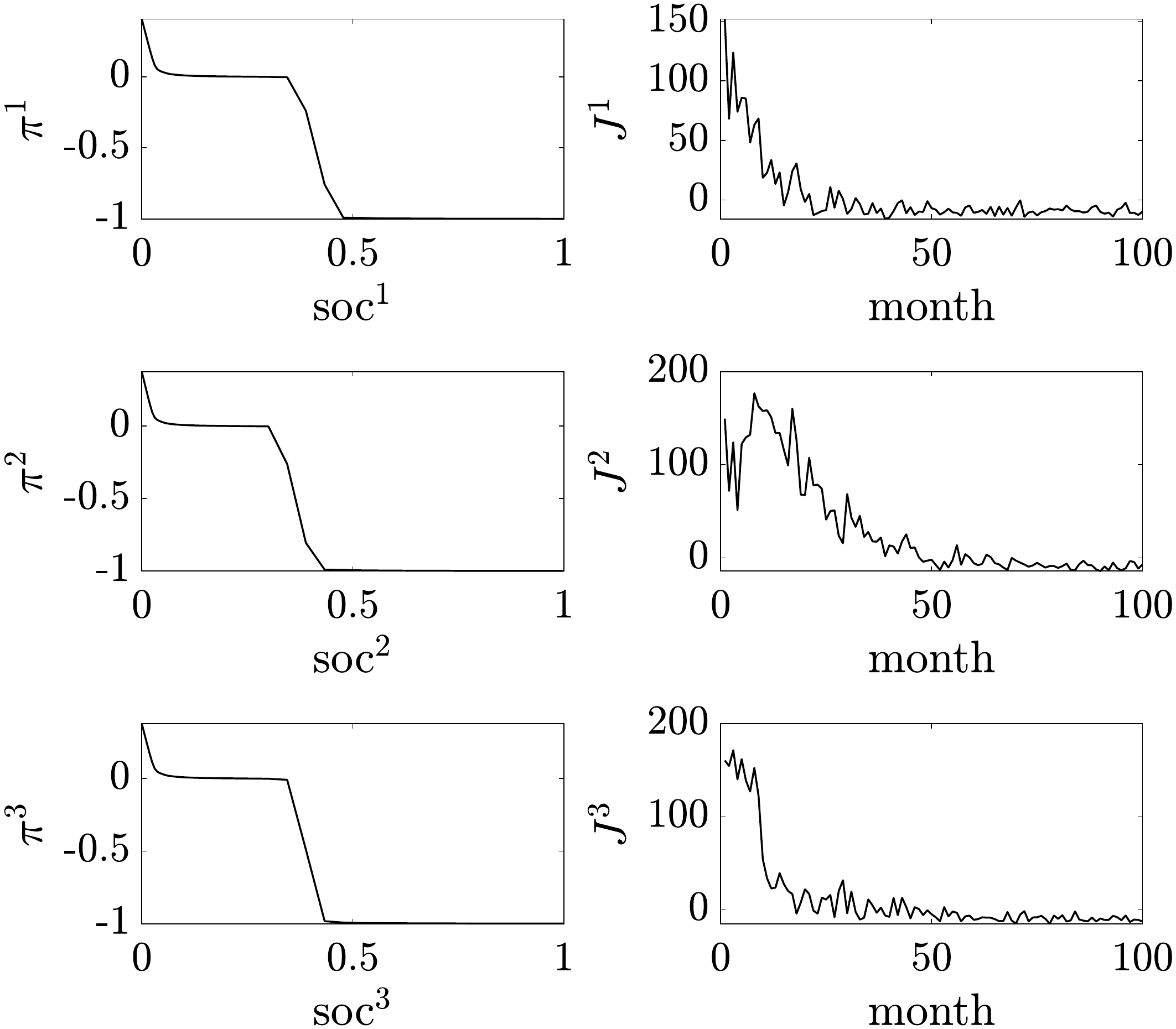}
\caption{(Left) The learned policy of each agent. (Right) The closed-loop performance of each agent.}
\label{fig:5}
\end{figure}

 Figure \ref{fig:5} (Left) shows the learned policy for each agent after the last RL-step. From the previous work (\cite{kordabad2021mpc}), we know that the linear economic stage cost often yields a (nearly) bang-bang structure optimal policy when the battery dynamics are stochastic and linear. This figure demonstrates the similar optimal policy consequence as expected.
Fig.\ref{fig:5} (Right) illustrates the improvement of the closed-loop performance $J^i$ for each agent during the learning.
\par
\begin{figure}[ht!]
\centering
\includegraphics[width=0.48\textwidth]{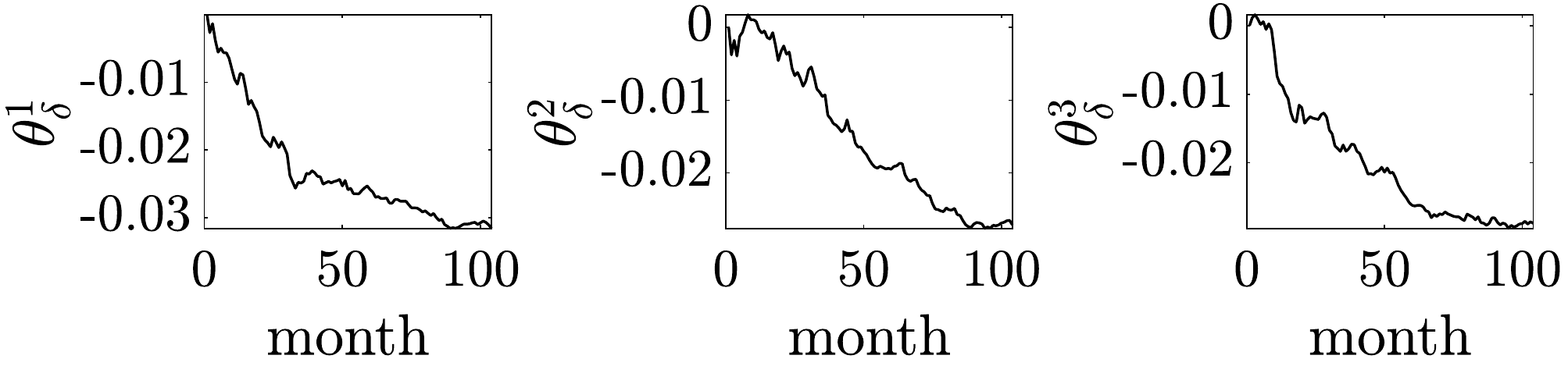}
\caption{Convergence of one of the policy parameters  $\theta^i_\delta$.}
\label{fig:7}
\end{figure}

 Fig.\ref{fig:7} illustrates the convergence of the parameter $\theta^i_\delta$. Note that there are $24$ parameters in this simulation, and we select $3$ representative parameters for the sake of brevity.

\section{CONCLUSION}\label{sec:conc}
In this paper, we propose an MPC-based RL approach to seek for an optimal policy for the multi-agent battery storage system. The objective is to minimize an economic cost considering the battery health using penalty for very low and high state of charge. We consider the production-demand uncertainty as well as the constraints for the peak power  exchanged with the main grid. We parametrize an MPC-scheme and use the deterministic policy gradient method to learn the optimal policy subject to the power peak constraints of the main grid. The simulation results prove the feasibility of the proposed method. For future works, we will use a decentralized learning on more comprehensive power systems, where the dynamics are more sophisticated and contain other uncertainties in the systems. 
\typeout{}
\bibliographystyle{IEEEtran}
\bibliography{mbattery}
\end{document}